\documentclass[10pt, conference]{IEEEtran}
\IEEEoverridecommandlockouts
\usepackage{cite}
\usepackage{amsmath,amssymb,amsfonts}
\usepackage{algorithmic}
\usepackage{graphicx}
\usepackage{textcomp}
\usepackage{xcolor}

\usepackage{graphicx}
\usepackage{amsmath}
\usepackage{algorithm}
\usepackage{paralist}
\usepackage{framed}
\usepackage{amsfonts}
\usepackage{multirow}
\usepackage{courier}
\usepackage{subcaption}
\usepackage{xcolor}
\definecolor{code-blue}{RGB}{8,0,255}
\definecolor{code-red}{RGB}{255,83,112}
\definecolor{code-purple}{RGB}{102,0,153}
\usepackage[american]{babel}
\usepackage{microtype}

\def\BibTeX{{\rm B\kern-.05em{\sc i\kern-.025em b}\kern-.08em
    T\kern-.1667em\lower.7ex\hbox{E}\kern-.125emX}}
\begin{document}

\title{Multi-Modal Attention Network Learning for Semantic Source Code Retrieval}

\author{\IEEEauthorblockN{Yao Wan\textsuperscript{*}$^{\dag\sharp}$, Jingdong Shu\textsuperscript{*}$^\dag$\thanks{\textsuperscript{*}Equal contribution.}, Yulei Sui$^\ddagger$, Guandong Xu$^\ddagger$, Zhou Zhao$^\text{\dag}$, Jian Wu$^\text{\dag}$ and Philip S. Yu$^{\dagger\diamond}$}
	\IEEEauthorblockA{$^\dag$College of Computer Science and Technology, Zhejiang University, Hangzhou, China\\
		$^\ddagger$School of Computer Science, University of Technology Sydney, Australia\\
		$^\dagger$Department of Computer Science, University of Illinois at Chicago, Illinois, USA\\
		$^\diamond$Institute for Data Science, Tsinghua University, Beijing, China\\
        $^\sharp$State Key Laboratory of Cognitive Intelligence, iFLYTEK, Hefei, China\\
		\{wanyao, jdshu, zhaozhou, wujian2000\}@zju.edu.cn, \{yulei.sui, guandong.xu\}@uts.edu.au, psyu@uic.edu}
}
	

\maketitle

\begin{abstract}
Code retrieval techniques and tools have been playing a key role in facilitating software developers to retrieve existing code fragments from available open-source repositories given a user query (e.g., a short natural language text describing the functionality for retrieving a particular code snippet). Despite the existing efforts in improving the effectiveness of code retrieval, there are still two main issues hindering them from being used to accurately retrieve satisfiable code fragments from large-scale repositories when answering complicated queries.
First, the existing approaches only consider shallow features of source code such as method names and code tokens, but ignoring structured features such as abstract syntax trees (ASTs) and control-flow graphs (CFGs) of source code, which contains rich and well-defined semantics of source code. Second, although the deep learning-based approach performs well on the representation of source code, it lacks the explainability, making it hard to interpret the retrieval results and almost impossible to understand which features of source code contribute more to the final results.

To tackle the two aforementioned issues, this paper proposes MMAN, a novel  \underline{M}ulti-\underline{M}odal \underline{A}ttention \underline{N}etwork for semantic source code retrieval. A comprehensive multi-modal representation is developed for representing unstructured and structured features of source code, with one LSTM for the sequential tokens of code, a Tree-LSTM for the AST of code and a GGNN (Gated Graph Neural Network) for the CFG of code. Furthermore, a multi-modal attention fusion layer is applied to assign  weights to different parts of each modality of source code and then integrate them into a single hybrid representation.
Comprehensive experiments and analysis on a large-scale real-world dataset show that our proposed model can accurately retrieve code snippets and outperforms the state-of-the-art methods.
\end{abstract}


\section{Introduction}
With the advent of immense source code repositories such as GitHub \cite{github} and StackOverflow \cite{stackoverflow}, it is gradually becoming a key software development activity for programmers to search existing code with the same functionality, and reuse as much of that code as possible \cite{reiss2009semantics}. 
The goal of code retrieval is to retrieve a particular code fragment from available open-source repositories given a user specification (e.g., a short text describing the functionality of the code fragment).
The key challenges of implementing such a code retrieval system lie in two folds: (a) a deep semantic understanding of the source code and (b) measuring the similarity of cross modalities (i.e., input natural language and source code).

\begin{figure}[!t]
	\centering
	\includegraphics[width=0.46\textwidth]{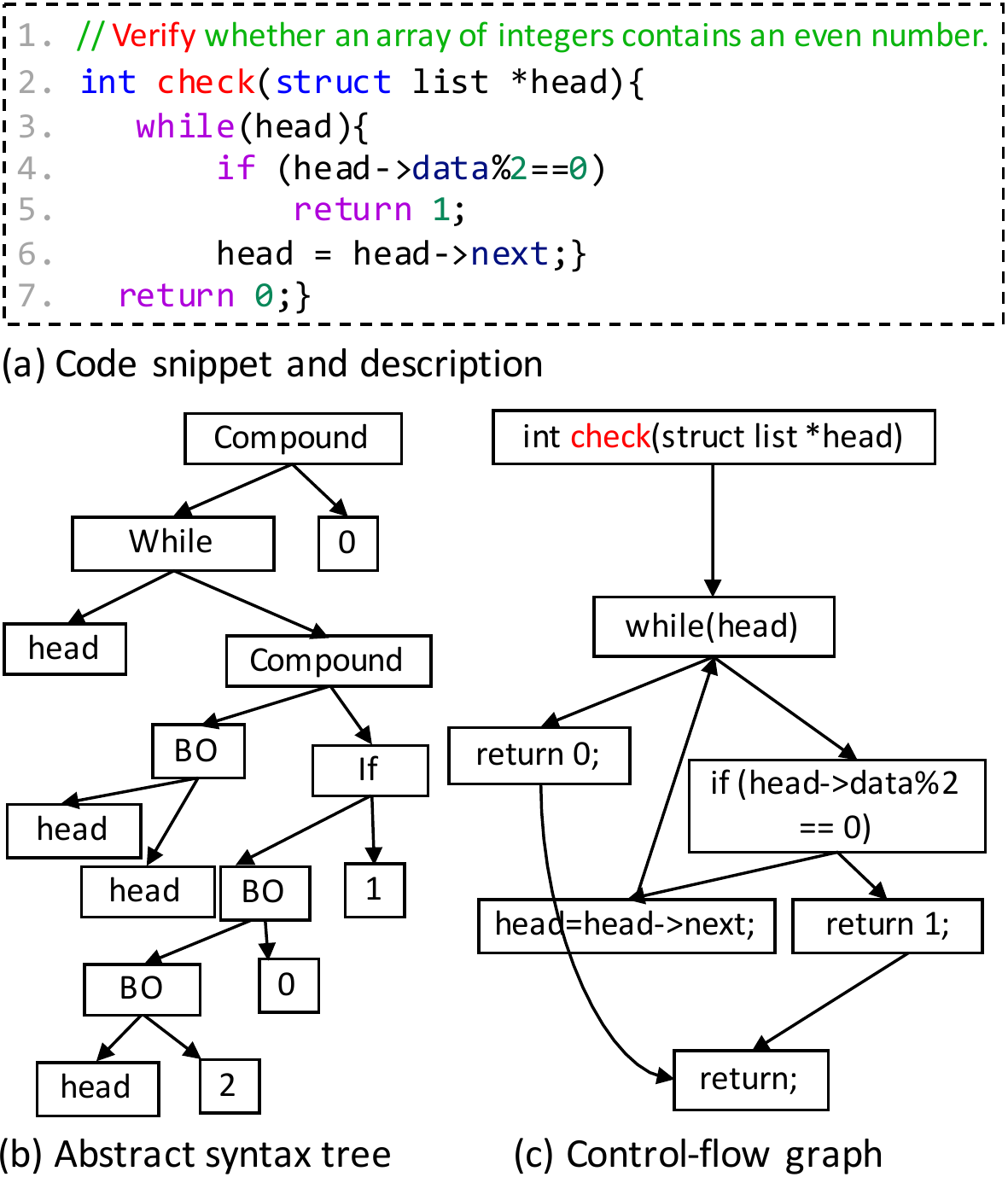}
	\caption{A motivating example to better illustrate our motivation. (a) A code snippet and its corresponding description. (b) The AST of the code snippet. (c) The control-flow graph of the code snippet.}
	\label{fig_cfg-ast}
	\vspace{-1em}
\end{figure}
\noindent\textbf{\textit{Existing Efforts and Limitations.}} 
Many existing efforts have been made towards searching the huge amount of available code resources for a natural language query, ranging from keyword matching \cite{lu2015query,lv2015codehow} to semantic retrieval \cite{reiss2009semantics, gu2018deep}. 
Lu et al., \cite{lu2015query} expanded a query with synonyms obtained from WordNet and then performed keyword matching of method signatures.
Lv et al., \cite{lv2015codehow} expanded the query with the APIs and considered the impact of both text similarity and potential APIs on code search.
Reiss et al., \cite{reiss2009semantics} developed a code retrieval system named Sourcerer, which learned the semantic representation of source code through a probabilistic topic model.
Inspired by the success of deep learning in computer vision and natural language processing tasks, deep learning has been applied to better represent source code for tasks such as clone detection \cite{white2016deep} and code summarization \cite{wan2018improving}.

To the best of our knowledge, Gu et. al., \cite{gu2018deep} is the first who applied deep learning network to the task of code retrieval, which captures the correlation between the semantic source code and natural language query in an intermediate semantic space. However, the approach still suffers from two major limitations: 
(a) \textit{Deep structured features of source code are often ignored}. The approach \cite{gu2018deep} captures the shallow source code information, including method name, code tokens and API sequence, missing the opportunity to capture the rich structure semantics of the code. 
(b) \textit{Lack of explainability}. The final results from a deep neural network is often hard to interpret since its internal working is always transparent to input data and different applications. This is also a common issue when applying deep learning models. For example, in \cite{gu2018deep}, the code and its natural language descriptions are projected into an intermediate semantic space and constrained by a ranking loss function. Although the semantic representation of code is learned, it is hard to infer which parts contribute more to the final result.

\noindent\textbf{\textit{Insights.}} These aforementioned limitations motivate us to design a model which learns a more comprehensive representation on source code as well as with the ability of explainability. From one hand, for limitation (a), apart from the \textit{tokens} of code, we also extract more features of code from its multiple views, such as \textit{abstract syntax tree} (AST) and \textit{control-flow graph} (CFG)~\footnote{The tree structure can also be seen as a special instance of graph with no circles and with each node having at most one parent node.}.
The AST and CFG are two types of intermediate code, one of which represents the hierarchical syntactic structure of a program, and the other represents the computation and control flow of a program \cite{aho1986compilers}. 
In this paper, we argue that aggregating complementary information from multiple views of source code can enrich its representation.
In this paper, we use the term view and modality interchangeably.
We call the approach of learning code representation from its multiple views/modalities as \textit{multi-modal learning}.
To address the limitation (b), since different modalities reflect different features of the source code. Therefore, each modality may not contribute equally to the final code representation. For a given modality, it consists of many elements (tokens, nodes in AST/CFG), weights are assigned to different elements via  representation learning. Therefore,  we can infer which part contributes more to the final result from the final representation, making  explainability possible. In this paper, we design an attention mechanism to integrate the multi-modal features into a single hybrid code representation. 

\noindent\textbf{\textit{A Motivating Example.} } We give an example in Figure \ref{fig_cfg-ast} to better illustrate our ideas. Figure \ref{fig_cfg-ast}(a) shows a simple C code example, which aims to verify whether an array of integers contains an even number. Figures \ref{fig_cfg-ast}(b) and (c) represent the corresponding AST and inter-procedural CFG of code in Figure \ref{fig_cfg-ast}(a), respectively.
From Figure \ref{fig_cfg-ast}(a), we can see that the semantics of the highlighted  three words \texttt{Verify}, \texttt{array}, \texttt{even} can be precisely captured by different code representations, e.g., plain text (for \texttt{check}), type-augmented AST (for \texttt{BinaryOperator}) and CFG (for \texttt{while}). 
These representations pay attention to different structure information of the code at different views, e.g., each node on AST represents a token and each node on CFG represents a statement. This shows the necessity of considering various modalities to better represent the source code.
It is necessary to represent a code from multiple views, especially from the structured information, since the orders of tokens and statements on the two views can be different depending on different code representations. 
For example, based on plain text, the token after ``\texttt{while}" in Figure \ref{fig_cfg-ast} (a) is ``\texttt{()}" and then followed by ``\texttt{head}". Differently, on AST, there will be two possible tokens following ``\texttt{Compound}", i.e., branch test ``\texttt{if}", ``\texttt{BinaryOperator}", as shown in Figure \ref{fig_cfg-ast} (b). Similarly, after the token ``\texttt{\}}" in the last statement at line 6, there will be no token left based on plain text. However, based on CFG, the next token is ``\texttt{while}" at the beginning of loop function based on CFG. 
From Figure \ref{fig_cfg-ast}, we can also observe that there exists an alignment relationship among the code snippet and it is description. For example, the keyword \texttt{Verify} should be closely connected to the word \texttt{check} in code. That means, on code retrieval, we can infer which part of the retrieved code contributes most to the input query words. This is very important to the model explainability. 

\noindent\textbf{\textit{Our Solution and Contributions.}} 
To tackle the two aforementioned issues, in this paper, we propose a novel model called \underline{M}ulti-\underline{M}odal \underline{A}ttention \underline{N}etwork (MMAN) for semantic source code retrieval. 
We not only consider the sequential features which have been studied in previous works (i.e., \textit{method name} and \textit{tokens}), but also the structure features (i.e., AST and CFG extracted from code). 
We explore a novel multi-modal neural network to effectively capture these multi-modal features simultaneously. In particular, we employ a LSTM \cite{hochreiter1997long} to represent the sequential tokens of code snippet, a Tree-LSTM \cite{tai2015improved} network to represent the abstract syntax tree (AST) and a gated graph neural network (GGNN) \cite{li2015gated} to represent the CFG. To overcome the explainability issue, we design an attention mechanism to assign different weights to different parts of each modality of source code, with the ability of explanation.
To summarize, the main contributions of this paper are as follows.
\begin{itemize}
	\item We propose a more comprehensive multi-modal representation method for source code, with one LSTM for the sequential content of source code, a Tree-LSTM for the AST of source code and a GGNN for the CFG of source code. Furthermore, a multi-modal fusion layer is applied to integrate these three representations.
	\item To the best of our knowledge, it is the first time that we propose an attention network to assign different weights to different parts of each modality of source code, providing an explainability of our deep multi-modal neural network for representation.
	\item To verify the effectiveness of our proposed model, we validate our proposed model on a real-world dataset crawled from GitHub, which consists of $28,527$ C code snippets. Comprehensive experiments and analysis show the effectiveness of our proposed model when compared with some state-of-the-art methods.
\end{itemize}

\noindent\textbf{\textit{Organization.}} The remainder of this paper is organized as follows. Section \ref{sec_relatedwork} highlights some works related to this paper. In Section \ref{sec_preliminaries}, we provide some background knowledge on multi-modal learning and attention mechanism. In Section \ref{sec_methodology}, we first give an overview our proposed framework and then present each module of our proposed framework in detail. Section \ref{sec_experiments} describes the dataset used in our experiment and shows the experimental results and analysis. Section \ref{sec_discussion} presents a discussion on our proposed model, including the strength as well as some threats to validity and limitations existing in our model. Finally, we conclude this paper and give some future research directions in Section \ref{sec_conclusion}.

\section{Related Work}\label{sec_relatedwork}
In this section, we briefly review the related studies from three perspectives, namely deep code representation, multi-modal learning and attention mechanism.

\subsection{Deep Code Representation}
With the successful development of deep learning, it has also become more and more prevalent for representing source code in the domain of software engineering research.
In \cite{mou2016convolutional}, Mou et al. learn distributed vector representations using tree-structured convolutional neural network (Tree-CNN) to represent snippets of code for program classification.
Similarly, Wan et al. \cite{wan2018improving} apply the tree-structured recurrent neural network (Tree-LSTM) to represent the AST of source code for the task of code summarization. 
Piech et al. \cite{piech2015learning} and Parisotto et al. \cite{parisotto2016neuro} learn distributed representations of source code input/output pairs and use them to guide program synthesis from examples.
In \cite{li2015gated}, Li et al. represent heap state as a graph and proposed a gated graph neural network to directly learn its representation to mathematically describe the shape of the heap. 
Maddison and Tarlow \cite{maddison2014structured} and other neural language models (e.g. LSTMs in Dam et al. \cite{dam2016deep}) describe context distributed representations while sequentially generating code. Ling et al. \cite{ling2016latent} and Allamanis et al. \cite{allamanis2015bimodal} combine the code-context distributed representation with distributed representations of other modalities (e.g., natural language) to synthesize code. 

One limitation of the above mentioned approaches is that these approaches ignore CFG of source code, which also conveys rich semantic information. Furthermore, no unified network is proposed to effectively fuse these multiple modalities. To mitigate this issue, this papers resort to propose a multi-modal network to learn a more comprehensive representation of source code. 

\subsection{Multi-Modal Learning}  
One prevalent direction in multi-modal learning is on joint representation which has been applied in many applications such as image captioning \cite{chen2018abstractive}, summarization~\cite{zhang2019multi}, visual questioning answering \cite{kim2018multimodal} and dialog system \cite{hori2019end}. 
In \cite{chen2018abstractive}, Chen et al. propose an attentional hierarchical neural network to summarize a text document and its accompanying images simultaneously. In \cite{zhang2019multi}, Zhang et al. propose a multi-modal (i.e., image and long description of product) generative adversarial network for product title refinement in mobile E-commerce.
In \cite{kim2018multimodal}, Kim et al. propose a dual attention network to capture a high-level abstraction of the full video content by learning the latent variables of the video input, i.e., frames and captions.
Similarly, in \cite{hori2019end}, Hori et al. answer questions about images using learned audio features, image features and video description, for the audio visual scene-aware dialog.
Another direction in multi-modal learning is cross-modal representation learning for information retrieval, which is similar to our task. Cross-modal representation learning aims to learn representation of each modality via project them into an intermediate semantic space with a constraint.
In \cite{carvalho2018cross}, Carvalho et al. propose a cross-modal retrieval model aligning visual and textual data (like pictures of dishes and their recipes) in a shared representation space for receipt retrieval.
In \cite{ma2015multimodal} Ma et al. propose a neural architecture for cross-modal retrieval, which combines one CNN for image representation and one CNN for calculating word-level, phrase-level and sentence-level matching scores between an image and a sentence.
\cite{cao2016deep,jiang2017deep}, the authors learn hash functions that map images and text in the original space into a Hamming space of binary codes, such that the similarity between the objects in the original space is preserved in the Hamming space. 

In this paper, we draw the insights from multi-modal learning, but not limit to it. We not only design a multi-modal neural network to represent the code, but also apply an attention mechanism to learn which part of code contributes more to the final semantic representation.

\subsection{Attention Mechanism} 
Attention mechanism has shown remarkable success in many artificial intelligence domains such as neural machine translation \cite{bahdanau2014neural}, image captioning \cite{you2016image}, image classification \cite{xiao2015application} and visual question answering \cite{lu2016hierarchical}. 
Attention mechanisms allow models to focus on necessary parts of visual or textual inputs at each step of a task. Visual attention models selectively pay attention to small regions in an image to extract core features as well as reduce the amount of information to process. A number of methods have recently adopted visual attention to benefit image classification \cite{mnih2014recurrent, stollenga2014deep}, image generation \cite{gregor2015draw}, image captioning \cite{xu2015show}, visual question answering \cite{yang2016stacked, xiong2016dynamic, shih2016look}, etc. 
On the other hand, textual attention mechanisms generally aim to find semantic or syntactic input-output alignments under an encoder-decoder framework, which is especially effective in handling long-term dependency. This approach has been successfully applied to various tasks including machine translation \cite{bahdanau2014neural}, text generation \cite{li2015hierarchical}, sentence summarization \cite{wan2018improving, rush2015neural}, and question answering \cite{kumar2016ask}. In \cite{lu2016hierarchical}, Lu et al. propose a co-attention learning framework to alternately learn the image attention and the question attention for visual question answering. In \cite{nam2017dual}, Nam et al. propose a multi-stage co-attention learning framework to refine the attentions based on memory of previous attentions.
In \cite{paulus2017deep}, Paulus et al. combine the inter- and intra-attention mechanism in a deep reinforcement learng setting to improve the performance abstractive text summarization.
In \cite{zhang2018self}, Zhang et al. introduce a self-attention mechanism into convolutional generative adversarial networks.

To the best of our knowledge, no study has attempted to learn multimodal attention models for the task of code retrieval.

\section{Preliminaries}\label{sec_preliminaries}
In this section, we first mathematically formalize the code retrieval problem using some basic notations and terminologies. We then present some background knowledge of multi-modal learning and attention mechanism.
\subsection{Problem Formulation}\label{sec_problem}
To start with, we introduce some basic notations. 
Suppose that we have a set $\mathcal{D}$ of $N$ code snippets, with corresponding descriptions, i.e., $\mathcal{D}=\{<x_1, d_1> , <x_2, d_1>, \ldots, <x_N, d_N> \}$. 
Each code snippet and description can be seen as a sequence of tokens. Let $x_i=(x_{i1}, x_{i2},\ldots, x_{i|x_i|})$ be a sequence of source code snippet, $d_i=(d_{i1}, d_{d2},\ldots, d_{i|d_i|})$ be a sequence of a description, where $|\cdot|$ denotes the length of a sequence. As we declared before, we represent the source code from three modalities (i.e., tokens, AST and CFG). We denote the semantic representation of code snippet $x_i$ as $\mathbf{x}_i=<\mathbf{x}_i^{tok}, \mathbf{x}_i^{ast}, \mathbf{x}_i^{cfg}>$, where $\mathbf{x}_i^{tok}$, $\mathbf{x}_i^{ast}$, $\mathbf{x}_i^{cfg}$ denote the representation for the three modalities, respectively. 

Since the code snippet and its description are heterogeneous, the goal of this paper is to train a model to learn their representation in an intermediate semantic space, simultaneously. Then, in the testing phase, the model can return a similarity vector of each candidate code snippet for a given query.

\subsection{Multi-Modal Learning}
\begin{figure}[ht]
	\centering
	\includegraphics[width=0.48\textwidth]{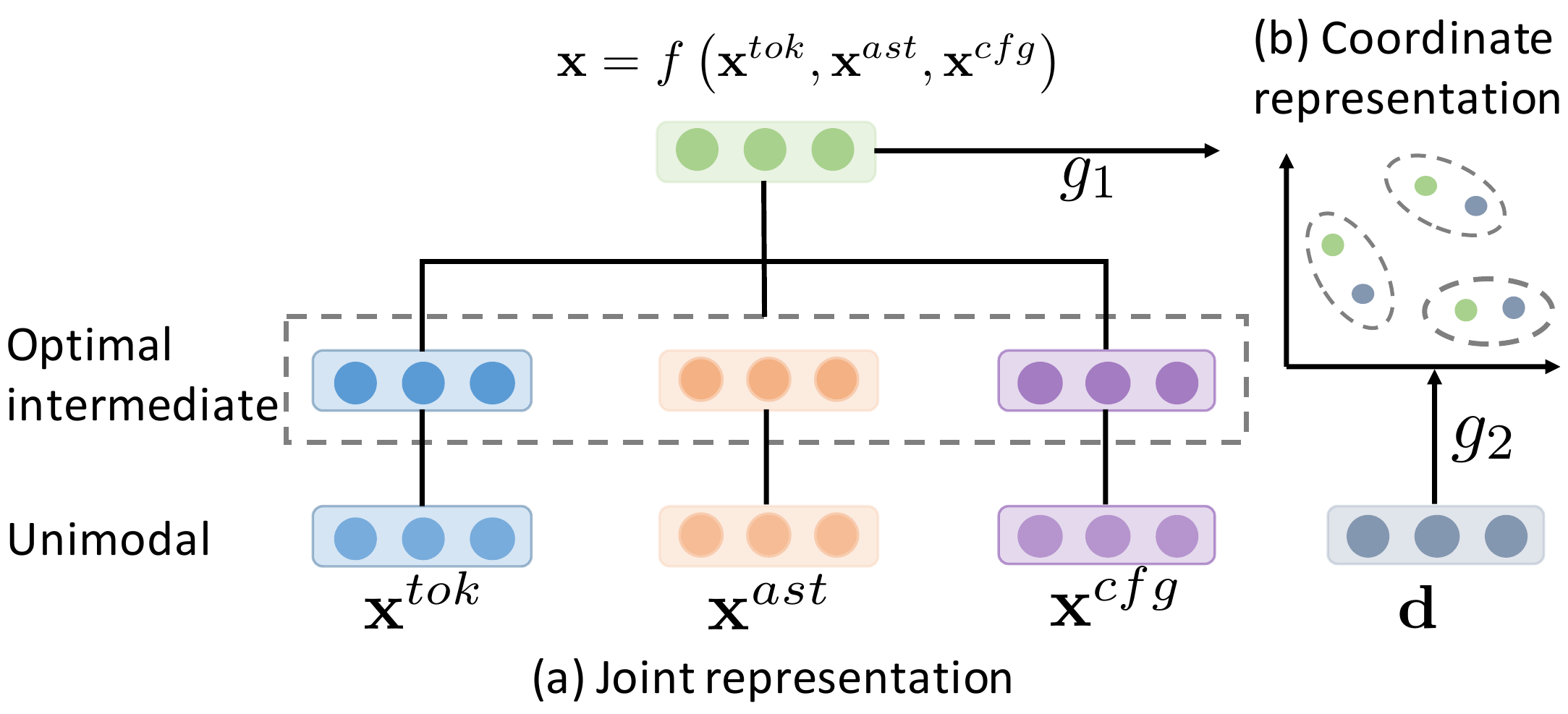}
	\caption{The difference and connection between \textit{joint} and \textit{coordinated} representations (adapted from~\cite{baltruvsaitis2019multimodal}).}
	\label{fig_multi-modal}
\end{figure}

Multi-modal learning aims to build models that can process and aggregate information from multiple modalities \cite{baltruvsaitis2019multimodal}. One important task of multi-modal learning is multi-modal representation learning, which is roughly categorized as two classes: \textit{joint} and \textit{coordinated}. Joint representations combine the unimodal signals into the same representation space, while coordinated representations process unimodal signals separately, but enforcing certain similarity constraints on them to bring them to what we term an intermediate semantic space. We introduce these two kinds of techniques in our problem setting. Figure \ref{fig_multi-modal} illustrates the difference and connection between of \textit{joint} and \textit{coordinated} representations.

For code snippet $\mathbf{x}$, we extract its multiple modalities such as $\mathbf{x}^{tok}$, $\mathbf{x}^{ast}$, $\mathbf{x}^{cfg}$. Since these modalities are complementary representation of a same code, we can apply the joint representation, which is formularized as follows:
\begin{equation}
	\mathbf{x}=f\left(\mathbf{x}^{tok}, \mathbf{x}^{ast}, \mathbf{x}^{cfg}\right),
\end{equation}
where the multimodal representation $\mathbf{x}$ is computed using function $f$ (e.g., a deep neural network) that relies on unimodal representations $\mathbf{x}^{tok}, \mathbf{x}^{ast}, \mathbf{x}^{cfg}$. 

While considering the code snippet $\mathbf{x}$ and description $\mathbf{d}$, since these two modalities are from different sources, it is desirable for us to apply coordinated representation for them \cite{gu2018deep}, which is defined as follows:
\begin{equation}
	g_1\left(\mathbf{x}\right) \sim g_2\left(\mathbf{d}\right),
\end{equation}
where each modality has a corresponding projection function ($g_1$ and $g_2$ above) that projects it into an intermediate semantic space with a similarity constraint/coordination on them. Examples of such coordination include minimizing cosine distance, or maximizing correlation. In this paper, the cosine similarity function is adopted.
\subsection{Attention Mechanism}
Attention networks learn functions that provide a weighting over inputs or internal features to steer the information visible to other parts of a network. 
To some extent, it is biologically motivated by the fact that our retina pays visual attention to different regions of an image or correlate words in one sentence. 
To date, many variants of attention mechanism have been evolved. 
From another perspective, the attention mechanism can be seen as the process of soft-addressing in a memory unit. The source, composed of key $\mathbf{k}$ and value $\mathbf{v}$, can be seen as the content of memory. Given an input query, the goal of attention mechanism is to return an attention value. Formally, we can define the attention value among query, key and value as follows.
\begin{equation}
	\alpha(\mathbf{q}, \mathbf{k}) = \operatorname{softmax}(g(\mathbf{q}, \mathbf{k})),
\end{equation}
where  $\mathbf{q}$ is the query and $\mathbf{k}$ is the key, $g$ is the attention score function which measures the similarity between query and key. Usually, the $g$ has many options, such as multi-layer perceptron \cite{bahdanau2014neural}, dot product \cite{luong2015effective} and scaled dot product \cite{vaswani2017attention}.
We call this kind of attention as inter-attention. However, there exists a condition that the query is the key itself. In this condition, we call it intra-attention (also well known as self-attention) \cite{cheng2016long,parikh2016decomposable}, which exhibits a better balance between ability to model long-range dependencies and computational and statistical efficiency. 
After obtaining the attention score, the final attended vector can be represented as the weighted sum of each value in the memory:
\begin{equation}
	\mathbf{v} = \sum_i \alpha_i(\mathbf{q}, \mathbf{k})\mathbf{v}_i,
\end{equation}
where $\mathbf{v}_i$ is the $i$-th value in the memory. 
In this paper, we adopt the inter-attention. Furthermore, the key is the hidden state of token or node in AST/CFG, and the value is the corresponding context vector (cf. Sec. \ref{sec_attention_fusion}).

\section{Multi-Modal Attention Network}\label{sec_methodology}
\subsection{An Overview}

\begin{figure}[ht]
	\centering
	\includegraphics[width=0.48\textwidth]{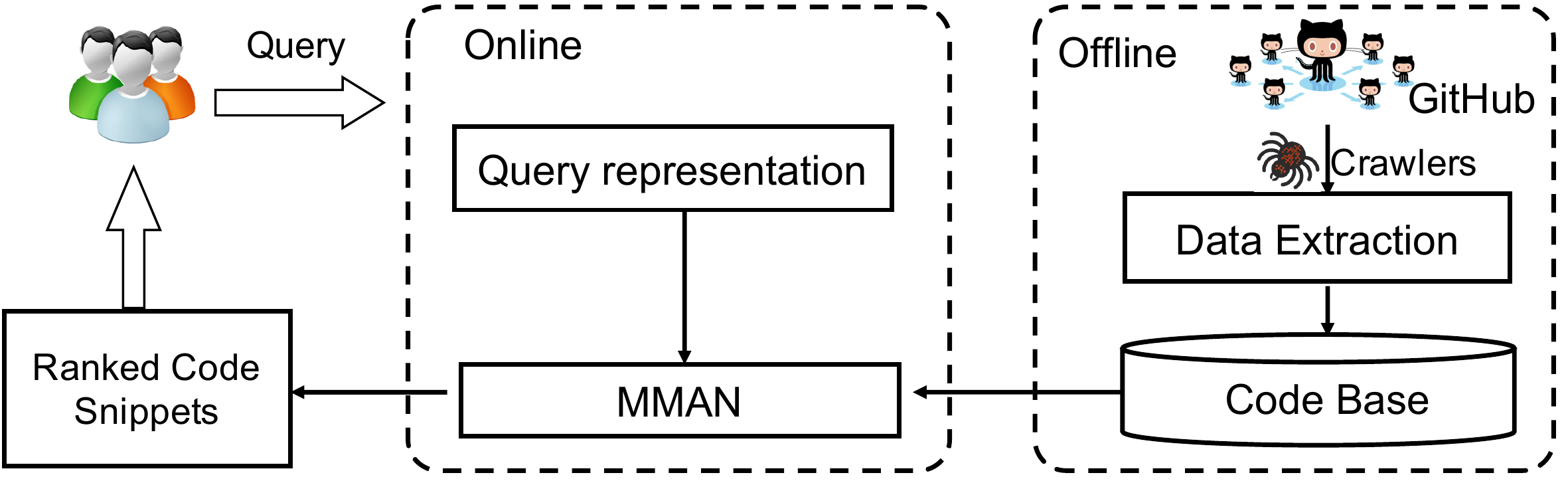}
	\caption{The workflow of MMAN.}
	\label{fig_workflow}
\end{figure}
In this section, we firstly give an overall workflow of how to get a trained model for code retrieval. Then we present an overview of the network architecture of our proposed MMAN model. 

Figure \ref{fig_workflow} shows the overall workflow of how to get a trained model, which consists of an offline training stage and an online retrieval stage. In the training stage, we prepare a large-scale corpus of annotated $<$code, description$>$ pairs. The annotated pairs are then fed into our proposed MMAN model for training. After training, we can get a trained retrieval network. Then, given a natural language query, related source code snippets can be retrieved by the trained network. 

\begin{figure*}[t]
	\centering
	\includegraphics[width=0.98\textwidth]{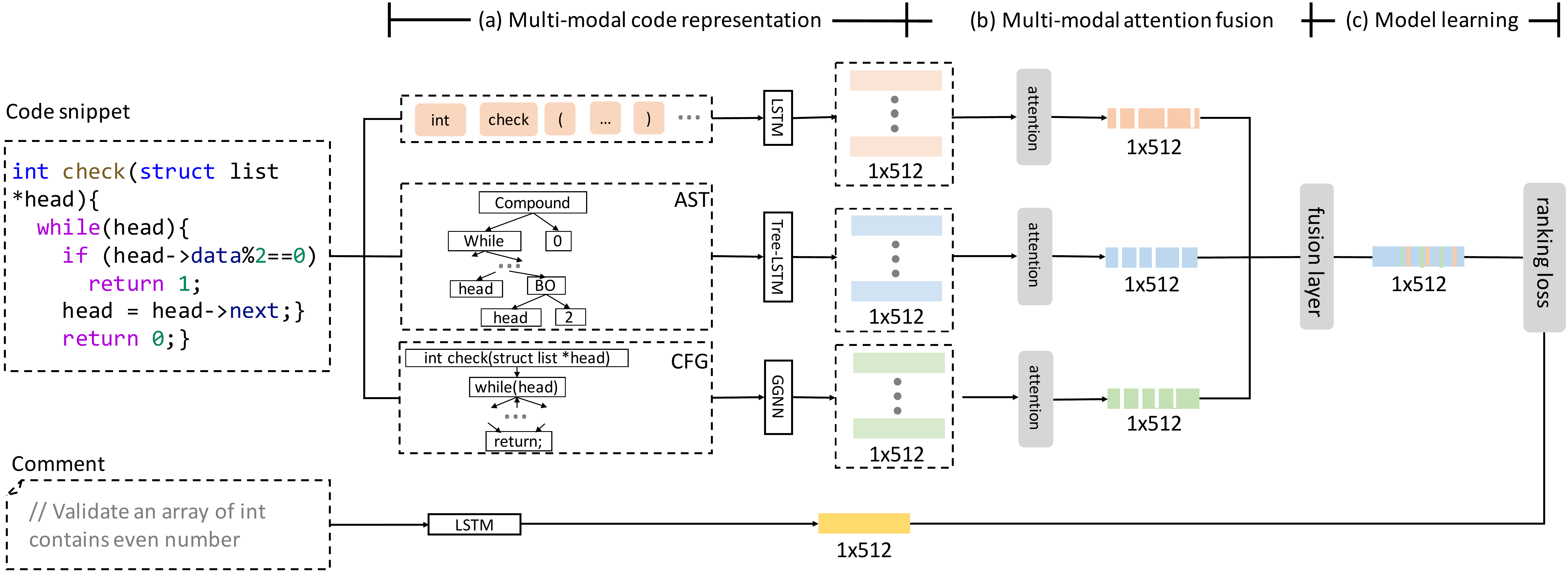}
	\caption{The network architecture of our proposed MMAN model. We first extract the $<$code, description$>$ pairs from training corpus. We then parse the code snippets into tree modalities, i.e., tokens, AST, CFG. Then the training samples are fed into the network as input. (a) Multi-modal code representation. We first learn the representation of each modality via LSTM, Tree-LSTM and GGNN, respectively. (b) Multi-modal attention fusion. We design an attention layer to assign different weight on different parts for each modality, and then fuse the attended vector into a single vector. (c) Model learning. We map the comment description representation and code representation into an intermediate semantic common space and design a ranking loss function to learn their similarities.}
	\label{fig_framework}
\end{figure*}

Figure \ref{fig_framework} is an overview of the network architecture of our proposed MMAN model. 
We split the framework into three submodules. (a) Multi-modal code representation (cf. Sec. \ref{sec_code_representation}). This module is used to represent the source code into a hidden space. (b) Multi-modal attention fusion (cf. Sec. \ref{sec_attention_fusion}). This attention module is designed to assign different weight on different parts for each modality, and then fuse the attended vector into a single vector. (c) Model learning (cf. Sec. \ref{sec_model_learning}). This attention module is designed to learn the comment description representation and code representation in a common space through a ranking loss function. We will elaborate each component in this framework in the following sections.

\subsection{Multi-Modal Code Representation}\label{sec_code_representation}
Different from previous methods that just utilize sequential tokens to represent code, we also consider the structure information of source code, in this section, we present a hybrid embedding approach for code representation. We apply a LSTM to represent the tokens of code, and a Tree-LSTM to represent the AST of code a GGNN to represent the CFG of code. 
\begin{figure}[t!]
	\centering
	\includegraphics[width=0.48\textwidth]{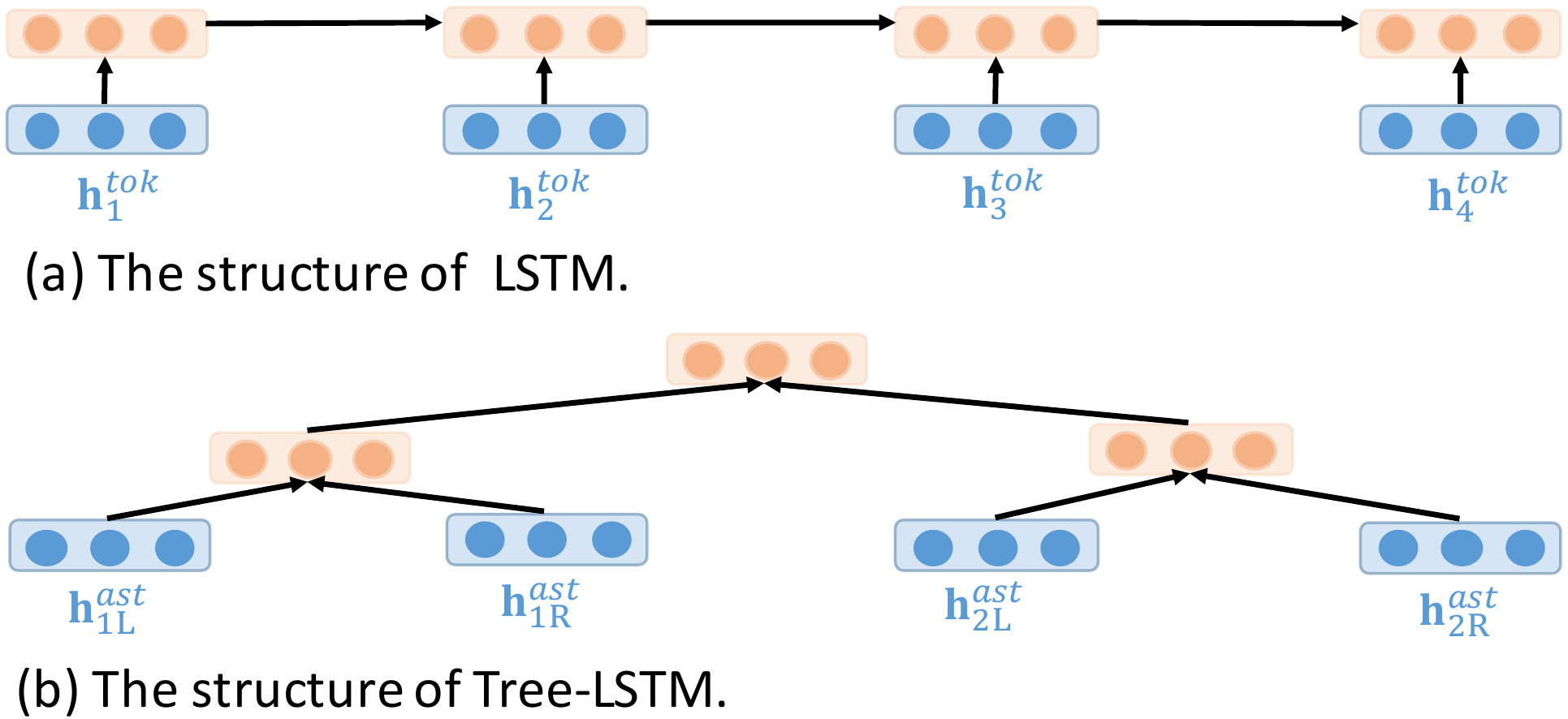}
	\caption{An illustratation of Tree-LSTM and LSTM.}
	\label{fig_lstm_treelstm}
\end{figure}
\subsubsection{Lexical Level - Tokens}
The key insight into lexical level representation of source code is that the comments are always extracted from the lexical of code, such as the method name, variable name and so on. In this paper, we apply LSTM network to represent the sequential tokens.
\begin{equation}
	\mathbf{h}_i^{tok}=\operatorname{LSTM}\left(\mathbf{h}_{i-1}^{tok}, w(x_i) \right), 
\end{equation}
where $i=1,\ldots, |x|$, $w$ is the word embedding layer to embed each word into a vector. 
The final hidden state $\mathbf{h}_{|x|}^{tok}$ of the last token of code is the token modality representation of $x$.
\subsubsection{Syntactic Level - AST}
We represent the syntactic level of source code from the aspect of AST embedding.
Similar to a traditional LSTM unit, we propose Tree-LSTM where the LSTM unit also contains an input gate, a memory cell and an output gate. However, different from a standard LSTM unit which only has one forget gate for its previous unit, a Tree-LSTM unit contains multiple forget gates. 
In particular, considering a node $N$ with the value $x_i$ in its one-hot encoding representation, and it has two children $N_L$ and $N_R$ , which are its left child and right child, respectively. The Tree-LSTM recursively computes the embedding for $N$ from the bottom up. Assume that the left child and the right child maintain the LSTM state $(\mathbf{h}_L, \mathbf{c}_L)$ and $(\mathbf{h}_R, \mathbf{c}_R)$, respectively.
Then the LSTM state $(h, c)$ of $N$ is computed as
\begin{equation}
	(\mathbf{h}_{i}^{ast}, \mathbf{c}_{i}^{ast})=\operatorname{LSTM}\left(\left(\left[\mathbf{h}_{iL}^{ast} ; \mathbf{h}_{iR}^{ast}\right],\left[\mathbf{c}_{iL}^{ast} ; \mathbf{c}_{iR}^{ast}\right]\right), w(x_i)\right),
\end{equation}
where $i=1,\ldots, |x|$ and $[\cdot; \cdot]$ denotes the concatenation of two vectors. Note that a node may lack one or both of its children. In this case, the encoder sets the LSTM state of the missing child to zero. In this paper, we adopt the hidden state of root node as the AST modality representation.
It's worth mentioning that when the tree is simply a chain, namely $N=1$, the Tree-LSTM reduces to the vallina LSTM. Figure \ref{fig_lstm_treelstm} shows the structure of LSTM and Tree-LSTM. 


\subsubsection{Syntactic Level - CFG}
\begin{figure}[t!]
	\centering
	\includegraphics[width=0.48\textwidth]{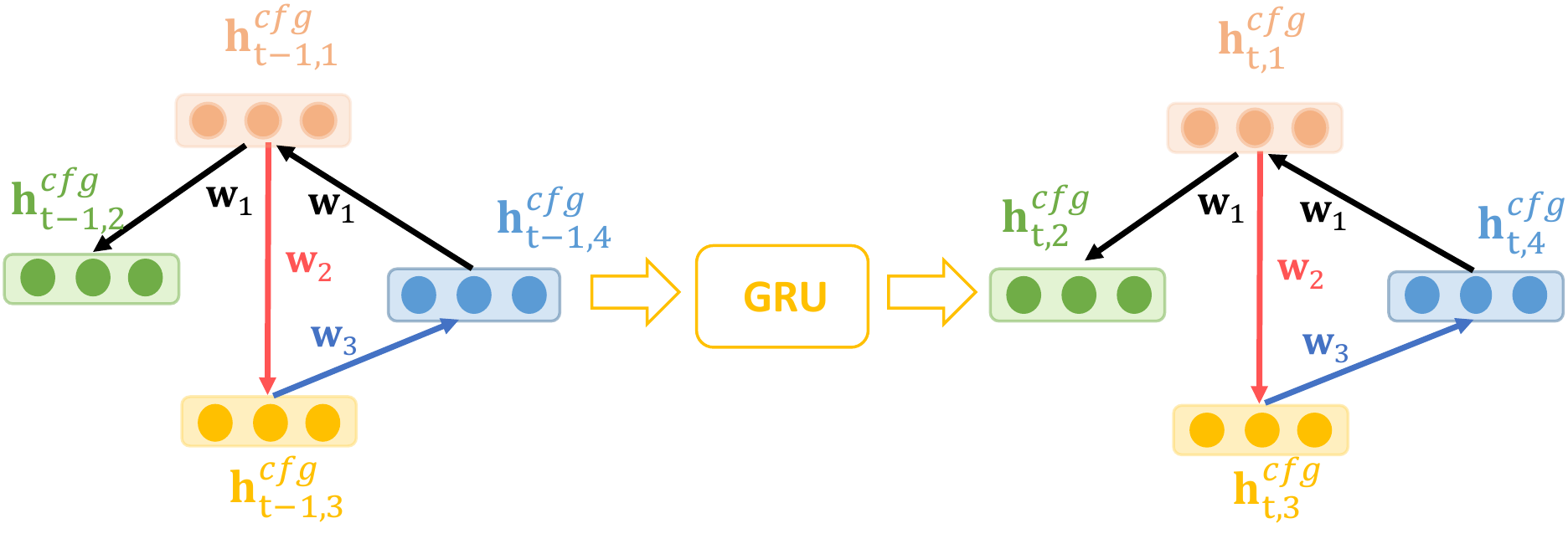}
	\caption{An illustration of GGNN.}
	\label{f_ggnn}
\end{figure}

As the CFG is a directed graph, we apply a gated graph neural network (GGNN) to represent the CFG, which is a neural network architecture developed for graph. We first define a graph as $\mathcal{G} = \{\mathcal{V},\mathcal{E}\}$, where $\mathcal{V}$ is a set of vertices $(v, \ell_v)$ and $\mathcal{E}$ is a set of edges $(v_i, v_j, \ell_e)$. $\ell_v$ and $\ell_e$ are labels of vertex and edge, respectively. In our code retrieval scenario, each vertex is the node of CFG, and each edge represents the control-flow of code, which has multiple types.
GGNN learns the graph representation directly through the following procedures: First, we initialize the hidden state for each vertex $v\in\mathcal{V}$ as $\mathbf{h}_{v,0}^{cfg} = w(\ell_v)$, where $w$ is the one-hot embedding function.
Then, for each round $t$, each vertex $v\in\mathcal{V}$ receives the vector $\mathbf{m}_{v, t+1}$, which is the ``message" aggregated from its neighbours. The vector $\mathbf{m}_{v, t+1}$ can be formulated as follows:
\begin{equation}
\boldsymbol{m}_{v,t+1}=\sum_{v' \in \mathcal{N}(v)} \mathbf{W}_{\ell_e} \mathbf{h}_{v',t},
\end{equation}
where $\mathcal{N}(v)$ are the neighbours of vertex $v$.
For round $t$, message from each neighbour is mapped into a shared space via $\mathbf{W}_{\ell_e}$.

For each vertex $v\in\mathcal{V}$, the GGNN update its hidden state with a forget gate. In this paper, we adopt the gated recurrent unit (GRU)~\cite{chung2014empirical} to update the hidden state of each vertex, which can be formulated as follows.
\begin{equation}
\mathbf{h}_{v,t+1}^{cfg} = \operatorname{GRU}(\mathbf{h}_{v,t}^{cfg}, \mathbf{m}_{v,t+1}).
\end{equation}
Finally, with $T$ rounds of iterations, we aggregate the hidden states of all vertices via summation to obtain the embedded representation of the CFG.
Figure \ref{f_ggnn} illustrates the structure of GGNN.

\subsection{Multi-Modal Attention Fusion}\label{sec_attention_fusion}
After we obtain the semantic representation of each modality, we need to fuse them into a single representation. As we declare before, for a unimodal, since it is composed of many elements, it is desirable to assign different weights to each element. 

\noindent\textbf{\textit{Token Attention.} }For tokens, not all tokens contribute equally to the final semantic representation of code snippet. Therefore, we introduce the attention mechanism on tokens to extract the ones that are more important to the representation of a sequence of code tokens. The attention score for tokens $\alpha^{tok}$ is calculated as follows:
\begin{equation}
    \alpha_{i}^{tok} = \frac{\exp (g^{tok}(f^{tok}(\mathbf{h}_i^{tok}), \mathbf{u}^{tok}))}{\sum_j \exp (g^{tok}(f^{tok}(\mathbf{h}_j^{tok}), \mathbf{u}^{tok}))},
\end{equation}
where $\mathbf{h}_i^{tok}$ represents the hidden state of $i$-th token in code. $f^{tok}$ denotes a linear layer and $g^{tok}$ is the dot-product operator. $\mathbf{u}^{tok}$ is the context vector of token modality, which can be seen as a high level representation of sequential tokens of code. The word context vector $\mathbf{u}^{tok}$ is randomly initialized and jointly learned during the training process.

\noindent\textbf{\textit{AST Attention. }} For the AST, not all nodes contribute equally to the final semantic representation of code snippet, indicating that different construct occurring in the source code (e.g., \texttt{if-condition-then}) should also be considered distinctly. Similar to Token attention, the attention score for AST nodes $\alpha^{ast}$ is calculated as follows:
\begin{equation}
    \alpha_{i}^{ast} = \frac{\exp (g^{ast}(f^{ast}(\mathbf{h}_i^{ast}), \mathbf{u}^{ast}))}{\sum_j \exp (g^{ast}(f^{ast}(\mathbf{h}_j^{ast}), \mathbf{u}^{ast}))},
\end{equation}
where $\mathbf{h}_i^{ast}$ represents the hidden state of $i$-th node in the AST. $f^{ast}$ denotes a linear layer and $g^{ast}$ is the dot-product operator. $\mathbf{u}^{ast}$ is the context vector of AST modality, which can be seen as a high level representation of AST nodes of code.

\noindent\textbf{\textit{CFG Attention. }} For the CFG, different statement in the source code should also be assigned different weight for the final representation. Therefore, we assign each CFG nodes with the weight $\alpha^{cfg}$ as:
\begin{equation}
	\alpha_{i}^{cfg} = \operatorname{sigmoid}(g^{cfg}(f^{cfg}(\mathbf{h}_i^{cfg}), \mathbf{u}^{cfg}),
\end{equation}
where $\mathbf{h}_i^{cfg}$ represents the hidden state of $i$-th node in the CFG. $f^{cfg}$ denotes a linear layer and $g^{cfg}$ is the dot-product operator. $\mathbf{u}^{cfg}$ is the context vector of CFG modality, which can be seen as a high level representation of CFG nodes of code.
It's worth mentioning that CFG attention weighted by sigmoid function achieves better performance than that by softmax function from the experimental results.

\noindent\textbf{\textit{Multi-Modal Fusion. }}We then integrate the multi-modal representation into a single representation via their corresponding attention score. We first concatenate them and then feed them into a one-layer liner network, which can be formularized as follows.
\begin{equation}
	\mathbf{x} = \mathbf{W}\left[\sum_{i} \alpha_{i}^{tok} \mathbf{h}_{i}^{tok}; \sum_{i} \alpha_{i}^{ast} \mathbf{h}_{i}^{ast}; \sum_{i} \alpha_{i}^{cfg} \mathbf{h}_{i}^{cfg}\right],
\end{equation} 
where $\mathbf{x}$ is the final semantic representation of code snippet $x$, $[\cdot; \cdot]$ is the concatenation operation and $\mathbf{W}$ is the attention weight for each modality.

\subsection{Description Representation}
In the training phase, the descriptions are extracted from the code comments, while in the testing phase, the description are regarded as the input queries. In this paper, we apply a vallina LSTM to represent the description.
\begin{equation}
	\mathbf{h}_i^{des}=\operatorname{LSTM}\left(\mathbf{h}_{i-1}^{des}, w(d_i) \right), 
\end{equation}
where $i=1,\ldots, |d|$ and $w$ is the word embedding layer to embed each word into a vector. The hidden state of last step $\mathbf{h}_{|d|}^{des}$ can be used as a vector representation of $d$.
\subsection{Model Learning}\label{sec_model_learning}
Now we present how to train the MMAN model to embed both code and descriptions into an intermediate semantic space with a similarity coordination. 
The basic assumption of this joint representation is that if a code snippet and a description have similar semantics, their embedded vectors should be close to each other.
In other words, given an arbitrary code snippet $x$ and an arbitrary description $d$, we want it to predict a high similarity if $d$ is a correct description of $x$, and a small similarity otherwise. In training phase, we construct each training instance as a triple $<x, d^+, d^->$: for each code snippet $x$, there is a positive description $d^+$ (a correct description of $x$) as well as a negative description (an incorrect description of $x$) $d^-$ randomly chosen from the pool of all $d^+$'s. When trained on the set of $<x, d^+, d^->$ triples, the MMAN predicts the cosine similarities of both $<x, d^+>$ and $<x, d^->$ pairs and minimizes the hinge range loss \cite{gu2018deep,kiros2014unifying}: 
\begin{equation}
	\mathcal{L}(\theta)=\sum_{<{x}, {d}^{+}, {d}^{-}>\in \mathcal{D}} \max (0, \beta-{sim} (\mathbf{x}, \mathbf{d}^{+})+{sim} (\mathbf{x}, \mathbf{d}^{-})),
\end{equation}
where $\theta$ denotes the model parameters, $\mathcal{D}$ denotes the training dataset, $sim$ denotes the similarity score between code and description $\beta$ is a small constant margin. $\mathbf{x}$, $\mathbf{d}^+$ and $\mathbf{d}^-$ are the embedded vectors of $x$, $d^+$ and $d^-$, respectively. In our experiments, we adopt the cosin similarity function (cf. \ref{sec_code_retrieval}) and set the fixed $\beta$ value to $0.05$. Intuitively, the ranking loss encourages the similarity between a code snippet and its correct description to go up, and the similarities between a code snippet and incorrect descriptions to go down.

\subsection{Code Retrieval}\label{sec_code_retrieval}
After the model is trained, we can deploy it online for service. Given a code base $\mathcal{X}$, for a given query $q$, the target is to rank all these code snippets by their similarities with query $q$. We first feed the code snippet $x$ into the multi-model representation module and feed the query $q$ as description into the LSTM module to obtain their corresponding representations, denoted as $\mathbf{x}$ and $\mathbf{q}$. Then we calculate the ranking score as follows:

%
\begin{equation}
	sim(x,q) = \cos (\mathbf{x}, \mathbf{q})=\frac{\mathbf{x}^{T} \mathbf{q}}{\|\mathbf{x}\|\|\mathbf{q}\|},
\end{equation}
where $\mathbf{x}$ and $\mathbf{q}$ are the vectors of code and a query, respectively. The higher the similarity, the more related the code is to the given query.

\section{Experiments and Analysis}\label{sec_experiments}
To evaluate our proposed approach, in this section, we conduct experiments to answer the following questions:
\begin{itemize}
	\item \textbf{\textit{RQ1.}} Does our proposed approach improve the performance of code retrieval when compared with some state-of-the-art approaches?
	\item \textbf{\textit{RQ2.}} What is the effectiveness and the contribution of each modality of source code, e.g., 
	sequential tokens, AST, CFG of source code for the final retrieval performance, and what about their combinations?
	\item \textbf{\textit{RQ3.}} What is the performance of our proposed model when varying the code length, code length, code AST node number, code CFG node number and comment length?
	\item \textbf{\textit{RQ4.}} What is the performance of our proposed attention mechanism?  
	What is the explainability of attention visualization?
\end{itemize}
We ask RQ1 to evaluate our deep learning-based model compared to some state-of-the-art baselines, which will be described in the following subsection. We ask RQ2 in order to evaluate the performance of each modality extracted from source code. We ask RQ3 to analyze the sensitivity of our proposed model when varying the code length, code AST node number, code CFG node number and comment length. We ask RQ4 to verify the explainability of our proposed attention mechanism. In the following subsections, we first describe the dataset, some evaluation metrics and the training details. Then, we introduce the baseline for RQ1. Finally, we report our results and analysis for four research questions.
\subsection{Dataset Collection}
\begin{figure}[!t]
	\centering
	\begin{subfigure}[b]{1.72in}
		\includegraphics[width=\textwidth]{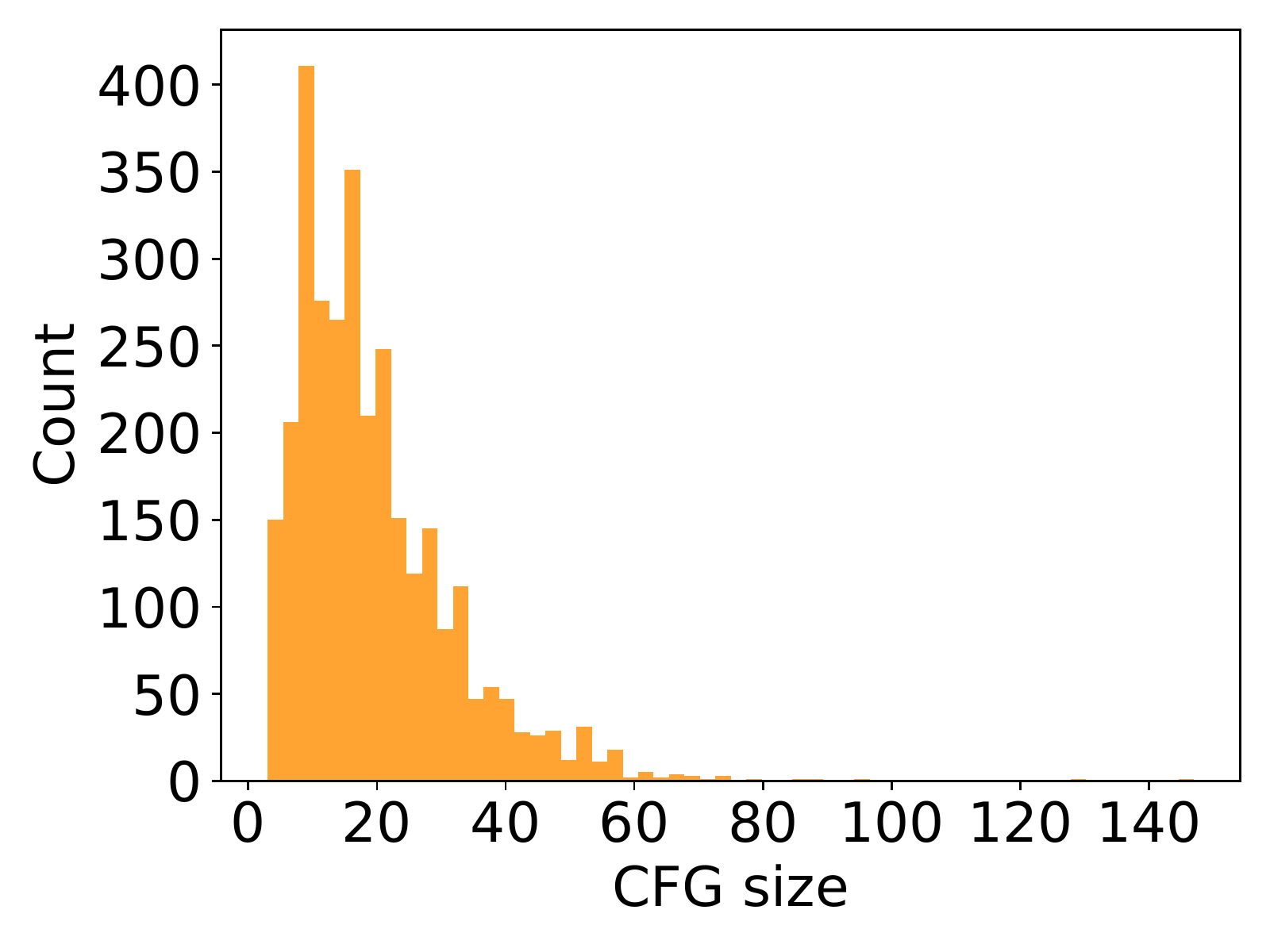}
		\caption{CFG node number.}
		\label{fig_cfg_node_num_distribution}
	\end{subfigure}
	\begin{subfigure}[b]{1.72in}
		\includegraphics[width=\textwidth]{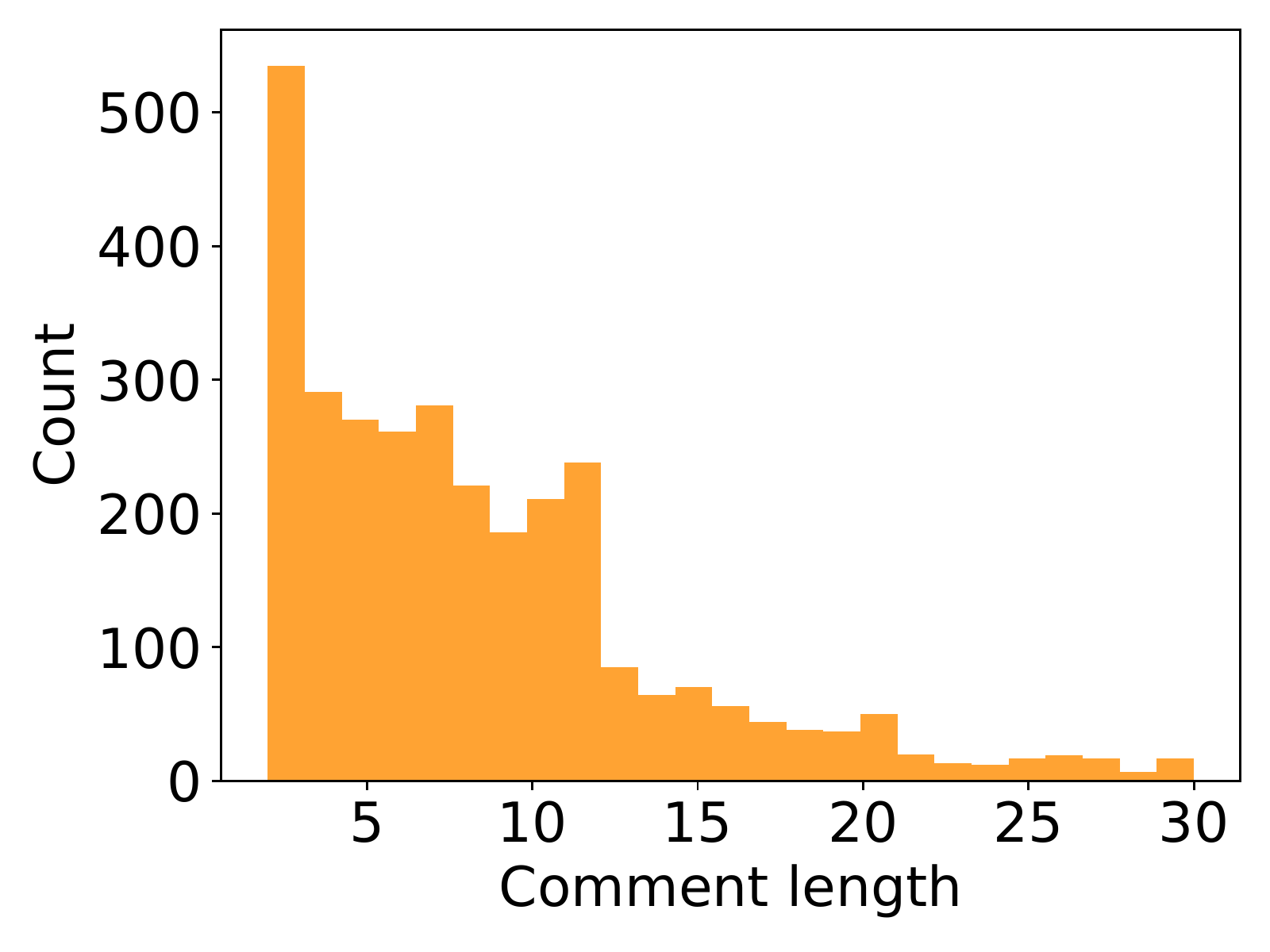}
		\caption{Comment length. }
		\label{fig_comment_length_distribution}
	\end{subfigure}
	\caption{The histogram of the dataset in our experiments. (a) CFG node number distribution. (b) Comment length distribution.}
	\label{fig_data_distribution}
\end{figure}
As described in Section \ref{sec_methodology}, our proposed model requires a large-scale training corpus that contains code snippets and their corresponding descriptions. Following but different from \cite{gu2018deep}, we evaluate the performance of our proposed model on a corpus of C code snippets, collected from GitHub (a popular open source projects hosting platform). 
Actually, we have considered the dataset released by \cite{gu2018deep}, while this dataset only contains the cleaned Java snippets without the raw data, unable to generate the CFG. Therefore, we resort to build a more complicated language C dataset, which may also provide more challenges and opportunities for our further research. 

To construct the codebase, we crawl all the C language repositories by its API\footnote{We crawled the GitHub in Oct., 2016, so the repositories in our database are created from August, 2008 to Oct., 2016.}. We then exclude the repositories whose stars number is smaller than 2. We select only the methods that have documentation comments from the crawled projects. Finally, we obtain a C corpus consisting of $28,527$ commented C methods.

Figure \ref{fig_data_distribution} shows the length distributions of code and comment on testing data. From Figure \ref{fig_cfg_node_num_distribution}, we can find that the lengths of most code snippets are located between 20 to 40. This was also observed in the quote in \cite{martin2009clean} ``Functions should hardly ever be 20 lines long". From Figure \ref{fig_comment_length_distribution}, we can notice that the lengths of nearly all the comments are smaller than 10. This also confirms the challenge for capturing the correlation between short text with its corresponding code snippet.

Having collected the corpus of commented code snippets, we extract the multi-modal code features and it's corresponding description, i.e., $<$method name, tokens, AST, CFG, description$>$, as follows:

\noindent\textbf{\textit{Method Name Extraction}.} For each C method, we extract its name and parse the name into a sequence of tokens according to camel case, and if it contains \texttt{\_}, we then tokenize it via \texttt{\_}.

\noindent\textbf{\textit{Token Extraction}.} To collect tokens from a C method, we tokenize the code by \texttt{\{. , " ; ) ( ! (space)\}}. After we tokenize function body, function name, we limit their max length as 100 and 50 respectively.

\noindent\textbf{\textit{AST Extraction}. }To construct the tree-structure of code, we parse C code into abstract syntax trees via an open source tool named Clang (\texttt{http://clang.llvm.org/}). For simplification, we transform the generated ASTs to binary trees by the following two steps which have been adopted in \cite{wan2018improving}: a) split nodes with more than 2 children, generate a new right child together with the old left child as its children, and then put all children except the leftmost as the children of this new node. Repeat this operation in a top-down way until only nodes with 0, 1, 2 children left; b) combine nodes with 1 child with its child.

\noindent\textbf{\textit{CFG Extraction}. } 
To construct the CFG of code, we first parse C function into CFG via an open source tool named SVF~\cite{sui2016svf} (\texttt{https://github.com/SVF-tools/SVF}), which has been widely used in value-flow analysis~\cite{sui2014detecting,sui2016demand}. We then remove nodes with same statement or no statement. For nodes with same statement, we retain the nodes which occur in the output of SVF first and remove their child nodes, and link children of their child nodes to them. For nodes without statement, we delete them and link their child nodes to their parent nodes. We set maximum size of CFG nodes as $512$.

\noindent\textbf{\textit{Description Extraction}.} To extract the documentation comment, we extract description via the regular expression \texttt{/**/}. We check the last sentence before every function and if it meets the condition that we have defined via regular expression, then we extract the description from the sentence.

We shuffle the dataset and split it into two parts, namely 27,527 samples for training and 1,000 samples for evaluation. 
It's worth mentioning a difference between our data processing and the one in \cite{gu2018deep}. In \cite{gu2018deep}, the proposed approach is verified on another isolated dataset to avoid the bias. Since the evaluation dataset doesn't have the ground truth, they manually labeled the searched results. We argue that this approach may introduce the human bias. Therefore, in our paper, we resort to the automatic evaluation. 



\subsection{Evaluation Metrics}
For automatic evaluation, we adopt two common metrics to measure the effectiveness of code retrieval, i.e., $\operatorname{SuccessRate}@k$ and Mean Reciprocal Rank (MRR), both of which have been widely used in the area of information retrieval. 
To measure the relevance of our search results, we use the success rate at rank $k$. The $\operatorname{SuccessRate}@k$ measures the percentage of queries for which more than one correct result could exist in the top $k$ ranked results \cite{kilickaya2016re,gu2018deep}, which is calculated as follows:
\begin{equation}
	\operatorname{SuccessRate}@ k = \left(\frac{1}{|Q|} \sum_{q=1}^{Q} \delta\left(\mathrm{FRank}_{q} \leq k\right)\right),
\end{equation}
where $Q$ is a set of queries, $\delta(\cdot)$ is a function which returns $1$ if the input is true and returns $0$ otherwise. $\operatorname{SuccessRate}@k$ is important because a better code search engine should allow developers to discover the needed code by inspecting fewer returned results. The higher the $\operatorname{SuccessRate}$ value is, the better the code search performance is.

We also use Mean Reciprocal Rank (MRR) to evaluate the ranking of our search results. The MRR \cite{lv2015codehow,gu2018deep} is the average of the reciprocal ranks of results of a set of queries $Q$. The reciprocal rank of a query is the inverse of the rank of the first hit result \cite{gu2018deep}. The $MRR$ is defined as
\begin{equation}
	\operatorname{MRR}=\frac{1}{|Q|}\sum_{q=1}^{|Q|}\frac{1}{\operatorname{FRank}_q},
\end{equation}
where $|Q|$ is the size of query set. The higher the MRR value is, the better the code retrieval performance is.

\subsection{Implementation Details}

To train our proposed model, we first randomize the training data and set the mini-batch size to 32. We build three separate vocabulary for code, comment and AST leaf node tokens with size 19,665, 9,628 and 50,004, respectively. For each batch, the code is padded with a special token \texttt{<PAD>} to the maximum length. All tokens in our dataset are converted to lower case. We set the word embedding size to 300. For LSTM and Tree-LSTM unit, we set the hidden size to be 512. For GGNN, we set the hidden size to 512 and set 5 rounds of iteration for GGNN. The margin $\beta$ is set to $0.05$. We update the parameters via Adam \cite{kingma2014adam} optimizer with the learning rate $0.0001$. To prevent over-fitting, we use dropout with $0.1$. 
All the models in this paper are trained for 100 epochs, taking about 41 hours.
All the experiments are implemented using the PyTorch 1.2 framework with Python 3.7, and the experiments were conducted on a computer with a Nvidia Quadro RTX 8000 GPU with 48 GB memory, running Ubuntu 18.04.


\subsection{Q1: Comparison with Baselines}
We compare our model with the following baseline methods:

\begin{itemize}
	\item CodeHow \cite{lv2015codehow} is a state-of-the-art code search engine proposed recently. It is an information retrieval based code search tool that incorporates an extended Boolean model and API matching. 
	
	\item DeepCS \cite{gu2018deep} is the state-of-the-art deep neural network based approach for code retrieval. It learns a unified vector representation of both source code and natural language queries.
	
	\item MMAN (Tok/AST/CFG)-w/o.Att. represents our proposed model on three modalities, without attention component for each modality. We also derive its variants on composition of these modalities. 
	
	\item MMAN (Tok/AST/CFG)-w.Att. represents our proposed model on three modalities, with attention component for each modality. We also derive its variants on combinations of these modalities. 
\end{itemize}

\noindent\textbf{\textit{Automatic Evaluation.}} We evaluate MMAN on the evaluation dataset, consisting of $1,000$ descriptions. In this automatic evaluation, we consider each description as an input query, and its corresponding code snippet as the ground truth. Table \ref{table_baseline} shows the overall performance of the three approaches, measured in terms of $SuccessRate@k$ and $MRR$. The columns $R@1$, $R@5$ and $R@10$ show the results of the average $SuccessRate@k$ over all queries when $k$ is $1$, $5$ and $10$, respectively. The column $MRR$ shows the $MRR$ values of the three approaches. 
From this table, we can draw the following conclusions: (a) Under all experimental settings, our MMAN (Tok+AST+CFG) method obtains higher performance in terms of both metrics consistently, which indicates better code retrieval performance. For the $R@k$, the improvements to DeepCS are 26.18\%, 16.06\% and 12.21\%, respectively. For the $MRR$, the improvement to DeepCS is 19.89\%. (b) Comparing the performance of two variants of our proposed model MMAN (Tok+AST+CFG)-w. or w/o.Att., we can observe that our designed attention mechanism indeed has a positive effect.



\begin{table}[!t]
	\centering
	\caption{Comparison of the overall performance between our model and baselines on automatic evaluation metrics. (Best scores are in boldface.)}
	\label{table_baseline}
	\begin{tabular}{|l|llll|}
		\hline
		Method& \textbf{R@1} &\textbf{R@5} & \textbf{R@10} & \textbf{MRR}\\
		\hline
		CodeHow      &0.262	&0.474	&0.543	&0.360      \\
		DeepCS      & 0.275	&0.498	&0.565	&0.377   \\
		\hline
		MMAN (Tok+AST+CFG)-w/o.Att.     & 0.243	&0.460	&0.517	&0.343\\
		MMAN (Tok+AST+CFG)-w.Att.     & \textbf{0.347}	&\textbf{0.578}	&\textbf{0.634}	&\textbf{0.452}\\
		\hline    
	\end{tabular}
\end{table}

\subsection{Q2: Effect of Each Modality}
To demonstrate the effectiveness of fusing multiple modalities, we have conducted experiments over different modality combinations to validate the effectiveness of fusing multiple modalities.
Table \ref{table_modality_effect} presents the performance of MMAN over various source combinations with and without attention.
From this table, we can observe that incorporating more modalities will achieve a better performance, which shows that there is a complementary rather than conflicting relationship among these modalities. In a sense, this is consensus with the old saying ``two heads are better than one". 
Comparing the performance of each modality comparison with or without attention, we also obtain that our designed attention mechanism has a positive effect on fusing these modalities together.

\begin{table}[!t]
	\centering
	\caption{Effect of each modality. (Best scores are in boldface.)}
	\label{table_modality_effect}
	\begin{tabular}{|l|llll|}
		\hline
		Method& \textbf{R@1} &\textbf{R@5} & \textbf{R@10} & \textbf{MRR}\\
		\hline
		MMAN (Tok)-w/o.Att.      &0.277	&0.468	&0.521	&0.365    \\
		MMAN (AST)-w/o.Att.      & 0.273	&0.458	&0.511	&0.358    \\
		MMAN (CFG)-w/o.Att.      & 0.110	&0.286	&0.363	&0.193    \\
		MMAN (Tok+AST+CFG)-w/o.Att.     & 0.243	&0.46	&0.517	&0.343 \\
		\hline
		MMAN (Tok)-w.Att.       &0.327	&0.562	&0.619	&0.432    \\
		MMAN (AST)-w.Att.       & 0.288	&0.496	&0.542	&0.379    \\
		MMAN (CFG)-w.Att.       & 0.119	&0.303	&0.387	&0.204    \\
		MMAN (Tok+AST+CFG)-w.Att.     & \textbf{0.347}	&\textbf{0.578}	&\textbf{0.634}	&\textbf{0.452}\\
		\hline    
	\end{tabular}
\end{table}

\subsection{Q3: Sensitivity Analysis}
\begin{figure}[!ht]
	\centering
	\begin{subfigure}[b]{1.72in}
		\includegraphics[width=\textwidth]{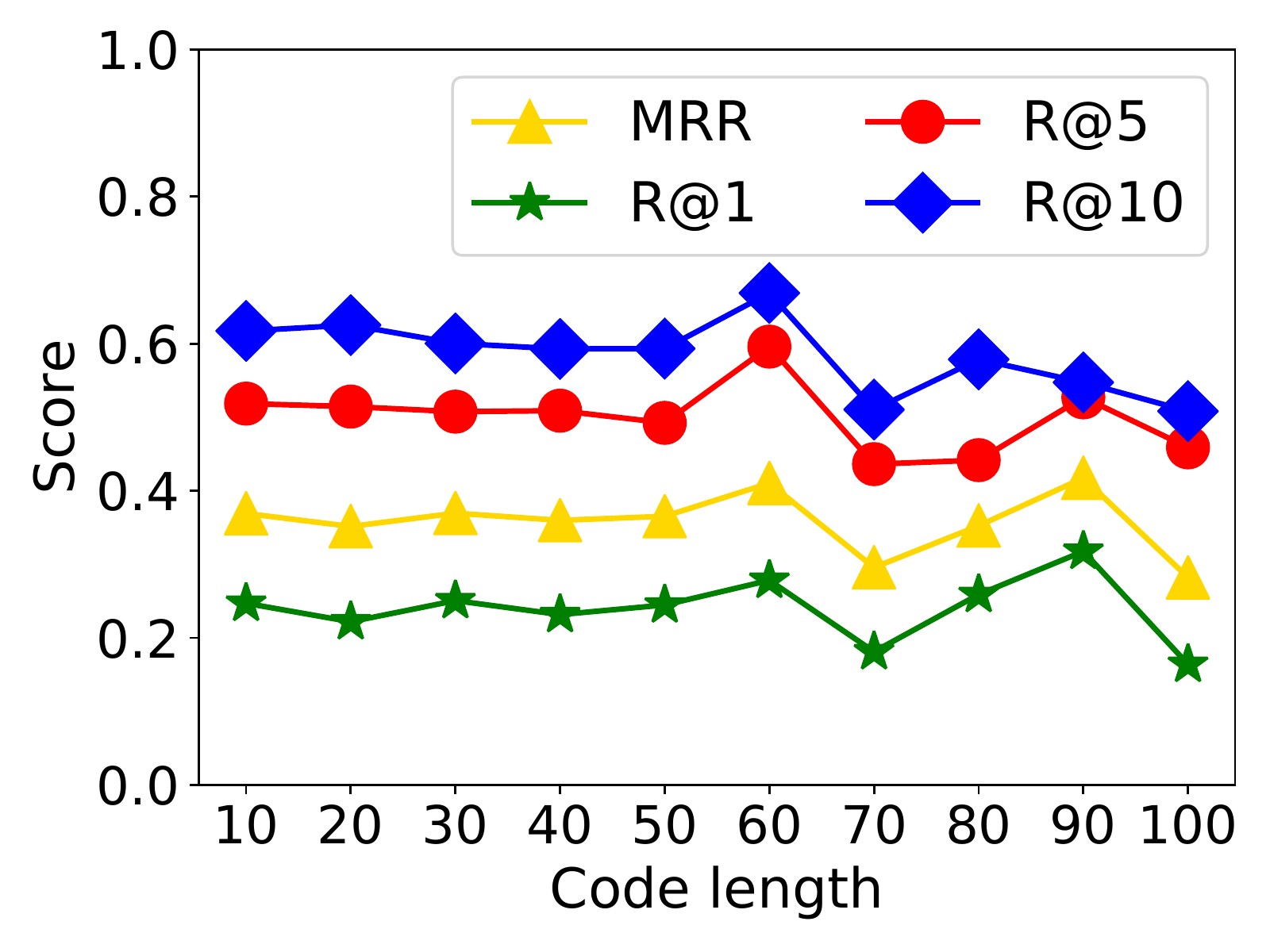}
		\caption{Code length.}
		\label{fig_varcode}
	\end{subfigure}
	\begin{subfigure}[b]{1.72in}
		\includegraphics[width=\textwidth]{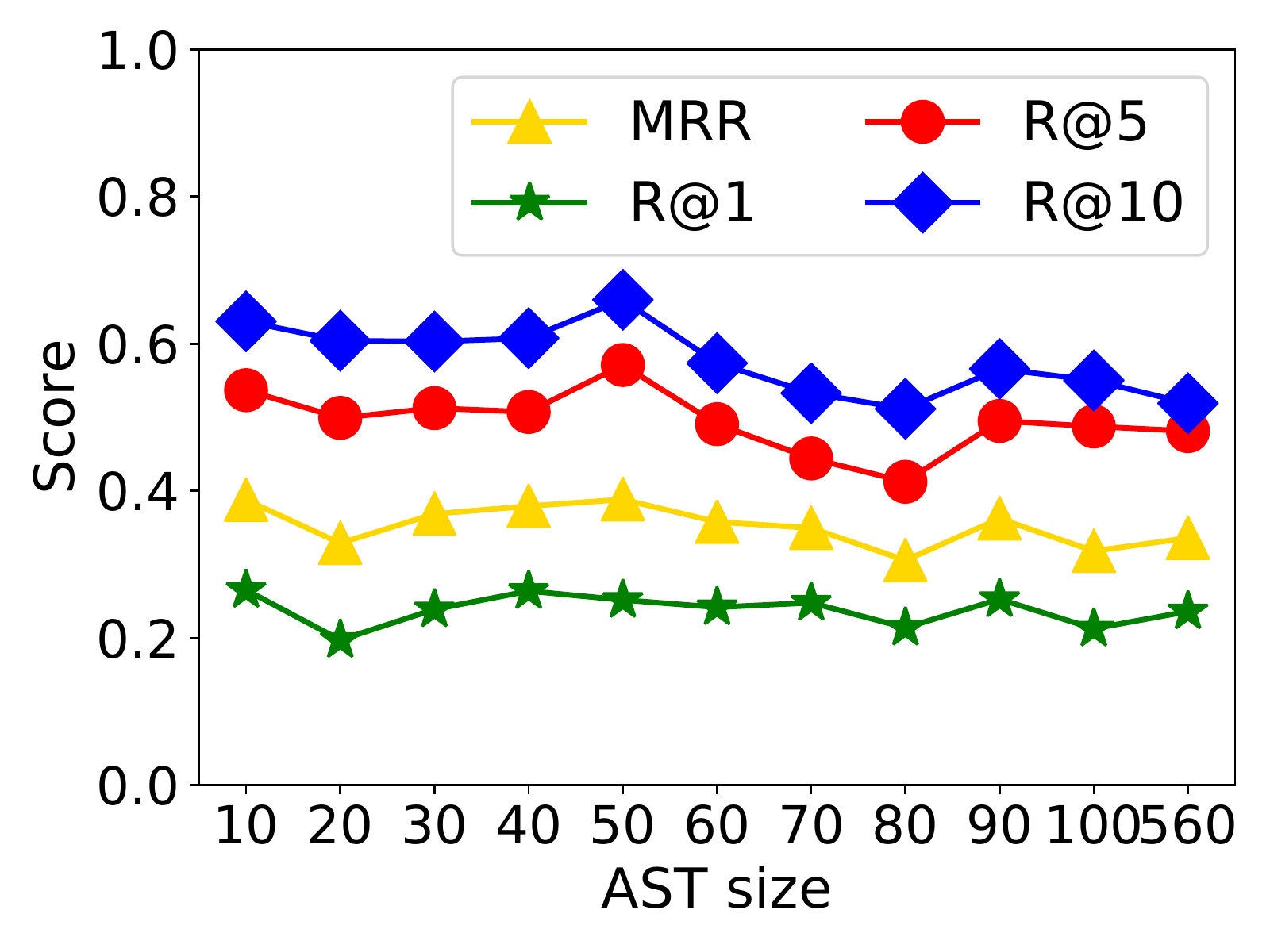}
		\caption{AST node number.}
		\label{fig_var_ast}
	\end{subfigure}
	\begin{subfigure}[b]{1.72in}
		\includegraphics[width=\textwidth]{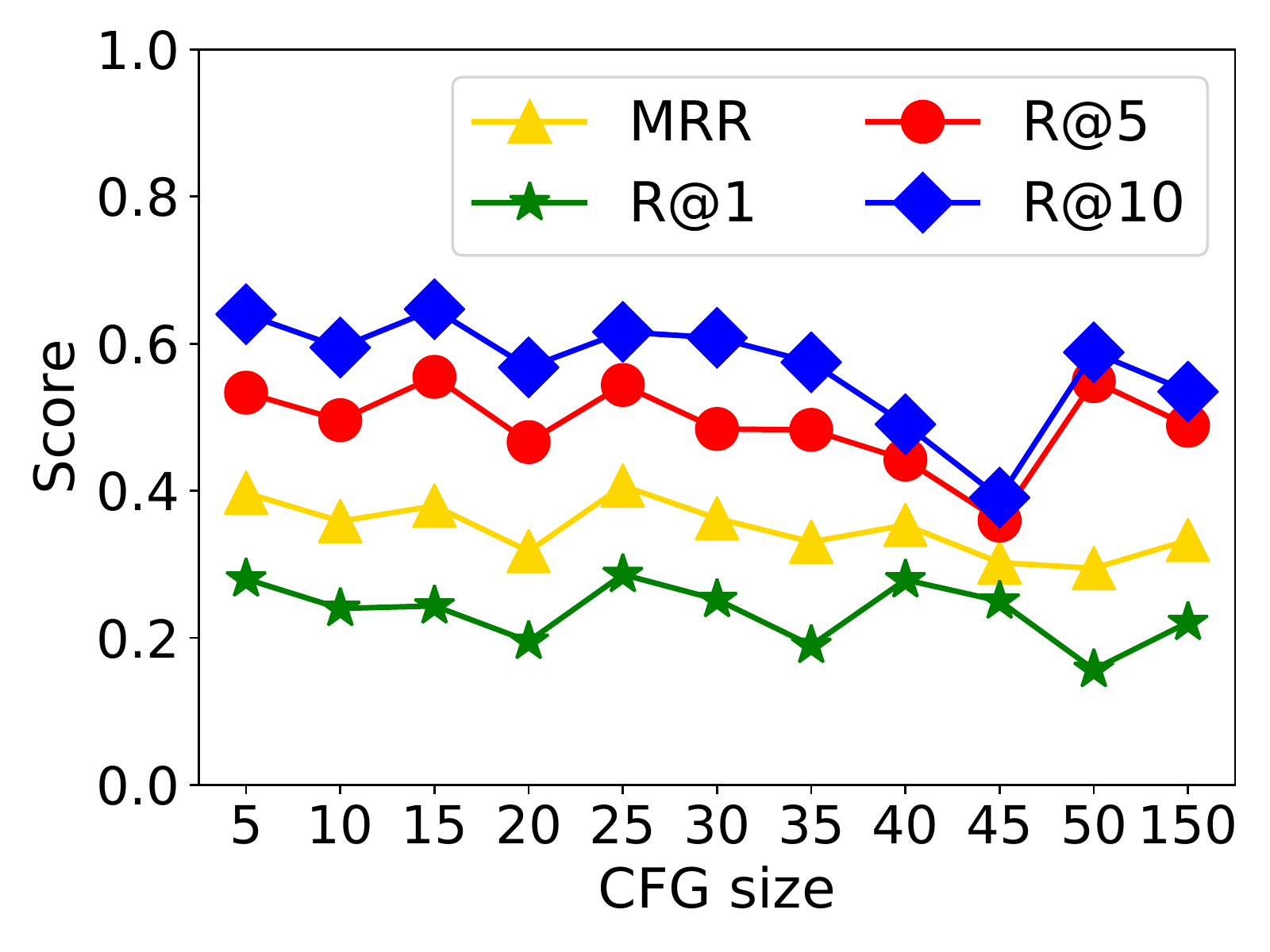}
		\caption{CFG node number.}
		\label{fig_var_cfg}
	\end{subfigure}
	\begin{subfigure}[b]{1.72in}
		\includegraphics[width=\textwidth]{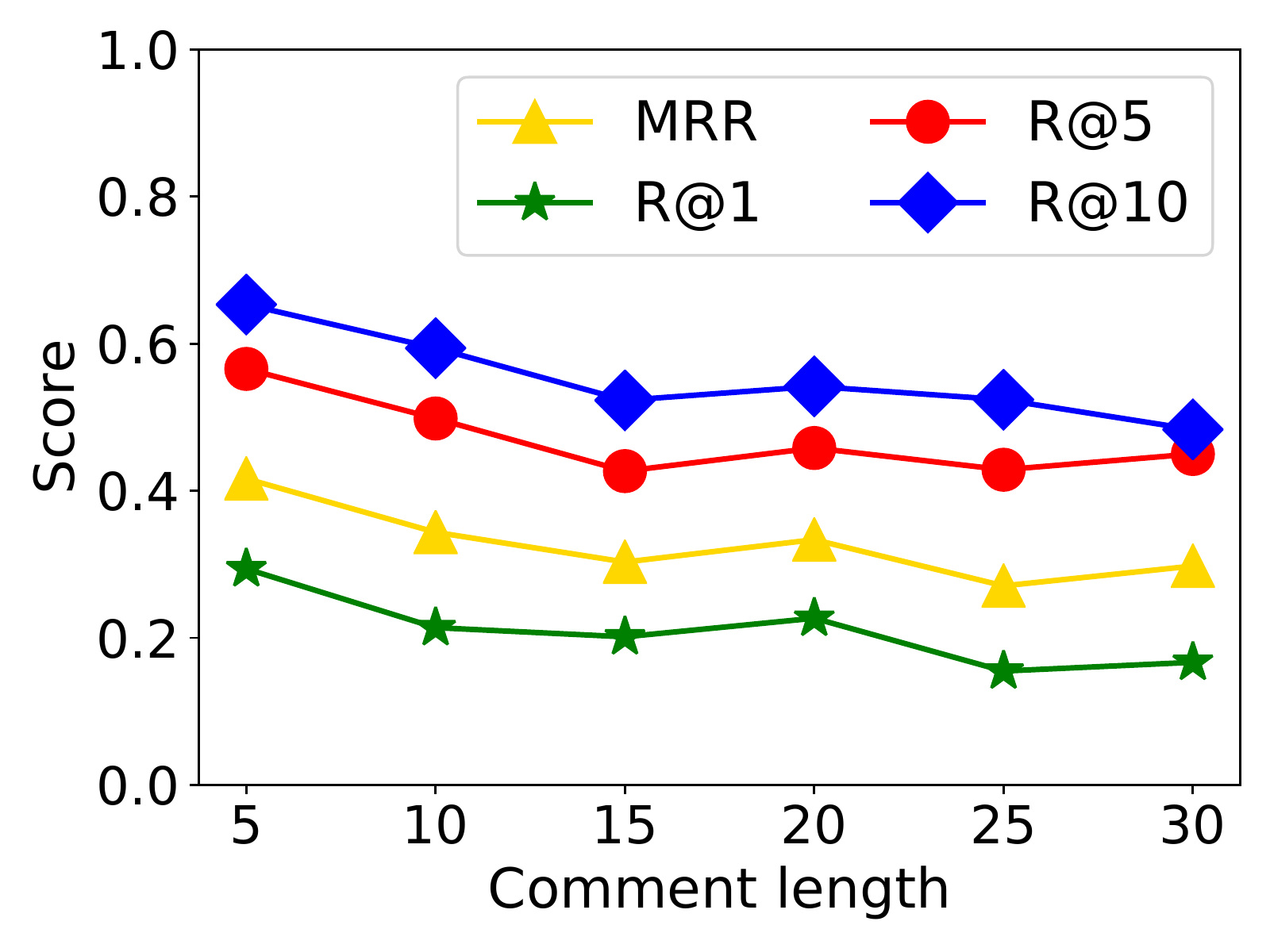}
		\caption{Comment length. }
		\label{fig_var_comment}
	\end{subfigure}
	\caption{Experimental results of our proposed method on different metrics w.r.t. varying code CFG node number and comment length.}
	\label{fig_parameter_analysis}
\end{figure}
To analyze the robustness of our proposed model, we study four parameters (i.e., code length, code AST node number, code CFG node number and comment length) which may have an effect on the code and comment representation.
Figure \ref{fig_varcode} shows the performance of our proposed method based on different evaluation metrics with varying code and comment lengths.
From Figures \ref{fig_varcode} (a)(b)(c), we can see that in most cases, our proposed model has a stable performance even though the code length or node number increases dramatically. We take this effect into account by attributing it to the hybrid representation we adopt in our model. 
From Figure \ref{fig_varcode}(d), we can see that the performance of our proposed model decreases on four metrics when the lengths of comments increase. This shows that increasing the length of a comment will increase the difficulty of comment understanding. Overall, it further verifies the robustness of multi-modal code representation.
\subsection{Q4: Qualitative Analysis and Visualization}
\begin{figure}[h!]
	\centering
	\includegraphics[width=0.48\textwidth]{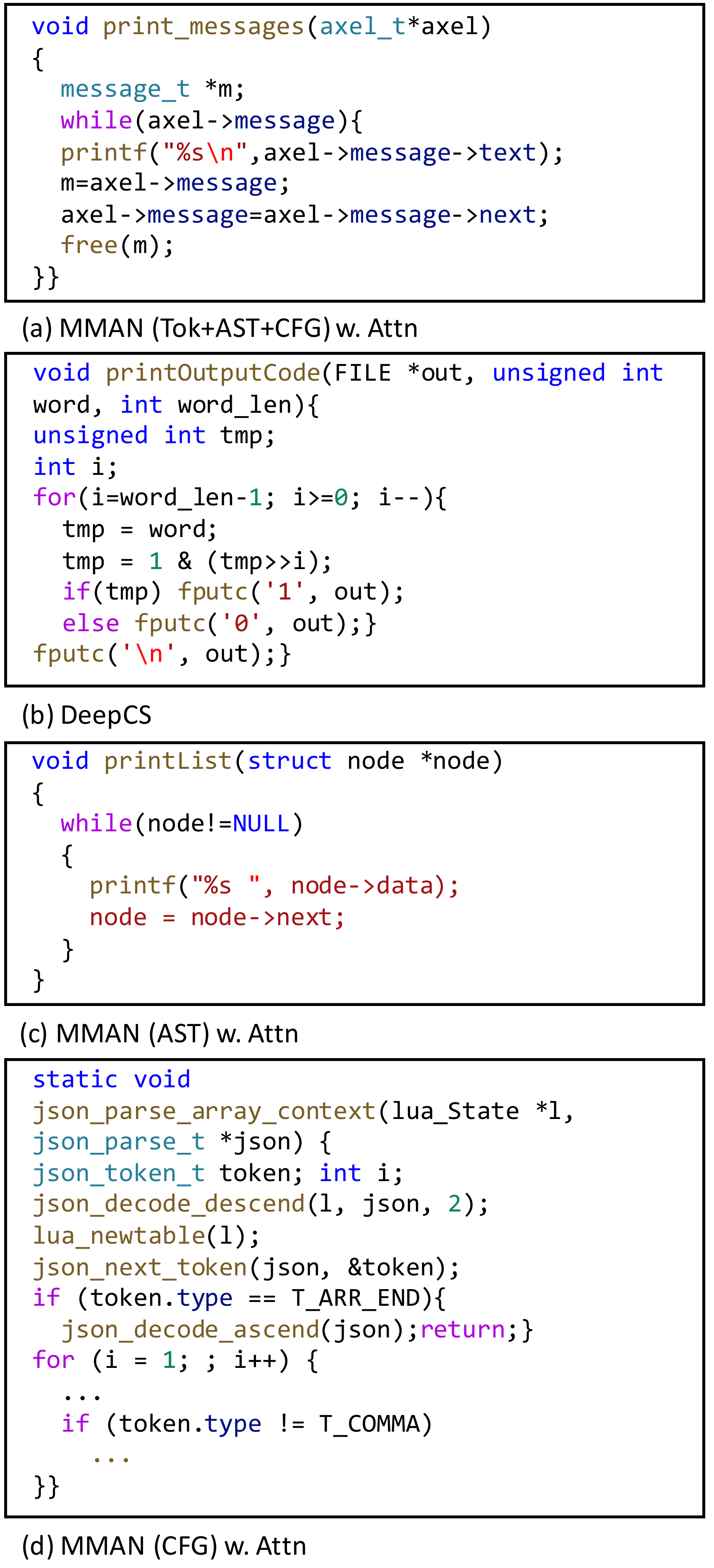}
	\caption{The retrieved first result of our proposed MMAN model with different modality combination and the DeepCS for query ``\textit{Print any message in the axel structure}"}
	\label{fig_cases}
\end{figure}
\begin{figure}[t!]
	\centering
	\includegraphics[width=0.48\textwidth]{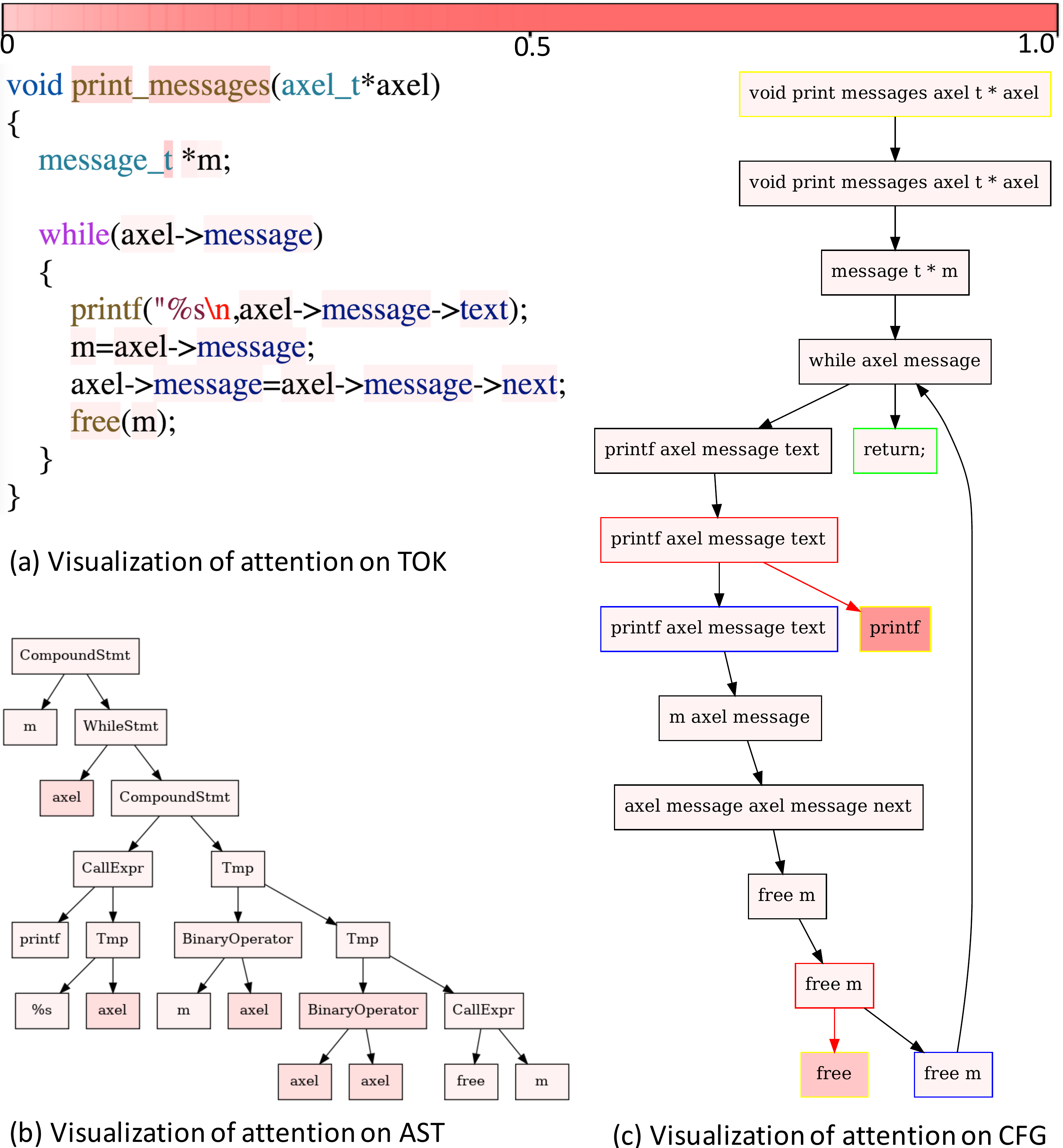}
	\caption{Attention visualization to interpret the contribution of each part of code w.r.t. different modalities.}
	\label{fig_attention_visualization}
\end{figure}
In the previous sections, we have shown the effectiveness of our proposed attention mechanism from the evaluation metrics. To gain some insights on the superiority of multi-modal representation and how the attention works, we show some cases retrieved by our model with different modality combinations (see Figure \ref{fig_cases}) and a visualization on the attention of each modality to interpret which part of code contributes more to the final result (see Figure \ref{fig_attention_visualization}).

Figure \ref{fig_cases} shows the first retrieved result of our proposed MMAN model with different modality combination and the DeepCS for query ``\textit{Print any message in the axel structure}". We can see that our proposed model can accurately retrieve the ground truth code snippets, when compared with the state-of-the-art DeepCS model (as shown in Figure \ref{fig_cases}b). 
When comparing Figure \ref{fig_cases}a with Figure \ref{fig_cases}c and Figure \ref{fig_cases}d, we can also clearly see the superiority of multi-modal approach for accurate code representation.

Figure \ref{fig_attention_visualization} visualizes the attention weights assigned to each part of code w.r.t. different modalities, for the sake of interpreting which part of code contributs more to the final result. From Figure \ref{fig_attention_visualization}a, we can see that the attention on tokens can really extract the important parts such as the function name \texttt{print\_message}. On the other hand, from Figure \ref{fig_attention_visualization}b, we can see the attention on AST assigns more weight on the leaf node (e.g., \texttt{axel}) as well as some operation nodes (e.g., \texttt{BinaryOperator}). Furthermore, from Figure \ref{fig_attention_visualization}c, we observe that the attention on CFG assigns more weight on the invoked function node (e.g., \texttt{prinf} and \texttt{free}). This can be illustrated by the fact that CFG can describe the control flow of a source code snippet.

\section{Discussion}\label{sec_discussion}
\subsection{Strength of MMAN}
We have identified three advantages of MMAN that may explain its effectiveness in code retrieval: (a) \textit{A more compresensive representation of source code from its multiple modalities.} The MMAN represents the source code snippet from its multiple modalities (i.e., tokens, AST and CFG) which contains complementary information for code representation. (b) \textit{An attention mechanism to assign different weights on different parts for each modality.} The MMAN contains an attention mechanism which can infer the contribution of each part of code to the final result, and through visualization, we can obtain a explainability for our deep learning based model, and  (c) \textit{A unified framework to learn the heterogeneous representation of source code and description in an intermediate semantic space.} The MMAN is an end-to-end neural network model with a unified architecture to learn the representation of source code and description in an intermediate semantic space.

\subsection{Threats to Validity and Limitations}\label{sec_threats}
Our proposed MMAN may suffer from two threats to validity and limitations. One threat to validity is on the evaluation metrics. We evaluate our approach using only two metrics, i.e., $SuccessRate@R$ and $MRR$, which are both standard evaluation metrics in information retrieval. We do not use precision at some cutoff ($Precision@k$), since the relevant results need to be labelled manually. We argue that this kind of approach will introduce the human bias. However, a human evaluation is also needed for the sake of fair comparison with DeepCS. 

Another threat to validity lies in the extensibility of our proposed approach. Our model needs to be trained on a large scale of corpus, which is collected from online platforms such as GitHub. Since the writing style of different programmers may differ greatly, lots of efforts will be put into this step. In this paper, we have defined many regular expressions to extract the samples that meets our condition, at the same time, many samples are filtered. Furthermore, the CFG can only be extracted from a whole program. Therefore, it's difficult to extend our multi-modal code representation model to some contexts where the CFG are unable to be extracted, such as many code snippets from StackOverflow.



\section{Conclusion and Future Work}\label{sec_conclusion}
In this paper, we have proposed a novel multi-modal neural network with attention machanism named MMAN for the task of code retrieval. 
Apart from considering the sequential features of source code such as \textit{tokens} and \textit{API sequences}, MMAN also considers the structure features of code such as AST and CFG to learn a more comprehensive semantic representation for code understanding. 
Furthermore, we put forward an attention mechanism to interpret the contribution of each part of code.
In addition, we proposed a unified framework to learn the representation of code representation and natural language query, simultaneously.
Our experimental study has shown that the proposed approach is effective and outperforms the state-of-the-art approaches. 

In our future work, we plan to conduct comprehensive experiments on other dataset of different language such as Java or Python, as well as human evaluation to further verify the effectiveness of our proposed approach. Furthermore, we believe that it's promising to explore the potentiality of multi-modal code representation on some other software engineering tasks such as code summarization and code clone detection.

\section*{Acknowledgment}
This paper is partially supported by the Subject of the Major Commissioned Project ``Research on China's Image in the Big Data" of Zhejiang Province's Social Science Planning Advantage Discipline ``Evaluation and Research on the Present Situation of China's Image" No. 16YSXK01ZD-2YB, the Zhejiang University Education Foundation under grants No. K18-511120-004, No. K17-511120-017, and No. K17-518051-021, the Major Scientific Project of Zhejiang Lab under grant No. 2018DG0ZX01, the National Natural Science Foundation of China under grant No. 61672453, the Key Laboratory of Medical Neurobiology of Zhejiang Province, and the Foundation of State Key Laboratory of Cognitive Intelligence (Grant No. COGOSC-20190002), iFLYTEK, P.R. China.
This work is also supported in part by NSF under grants III-1526499, III-1763325, III-1909323, SaTC-1930941, CNS-1626432, Australian Research Council Linkage Project under LP170100891, and Australian Research Grant DE170101081.

\ifCLASSOPTIONcaptionsoff
\newpage
\fi



%

\bibliographystyle{IEEEtran}
\bibliography{ref}

\end{document}